# OPTICAL CHARACTERISATION OF MOVPE GROWN VERTICALLY CORRELATED InAs/GaAs QUANTUM DOTS

*P. Hazdra[1], J. Voves[1], J. Oswald[2], K. Kuldová[2], A. Hospodková[2], E. Hulicius[2] and J. Pangrác[2]*

[1] Department of Microelectronics, Czech Technical University in Prague, Technická 2, 16627 Prague 6, Czech Republic
[2] Institute of Physics, Academy of Sciences of the Czech Republic, Cukrovarnická 10, 16253 Prague 6, Czech Republic

## ABSTRACT

Structures with self-organised InAs quantum dots in a GaAs matrix were grown by the low-pressure metal-organic vapour phase epitaxy (LP-MOVPE) technique. Photoluminescence in combination with photomodulated reflectance spectroscopy were used as the main characterisation methods for the growth optimisation. Results show that photoreflectance spectroscopy is an excellent tool for characterisation of QD structures wetting layers (thickness and composition) and for identification of spacers in vertically stacked QDs structures.

## 1. INTRODUCTION

Self-assembled InAs/GaAs quantum dots (QDs) are a subject of intense research due to their unique atomic-like properties and potential device applications. A part of the research activities is driven by the idea to bring the emission wavelength of the QDs up to the values of 1.3 µm and 1.5 µm important for telecommunications. This will allow to produce cheap and reliable laser structures on GaAs substrate. The performance of the QD-based light emitting structures is determined by the dot size, by the dot density and by the difference between the transition energy of the first excited state and that of the ground state ($\Delta E_{g-e}$). Obviously, when improving the wavelength, the values of these two crucial parameters should be kept as high as possible.

The emission wavelength of the QDs can be changed by changing their size and the composition of the dots and of the barriers [1]. The latter determines the lattice constants and the related strain, and the conduction- and valence-band offsets (heights of the barriers). Another way to achieve higher values of the wavelength consists in using vertically stacked multi-layer QD structures [2]. During the growth, the QDs form preferentially just above those in the previous layer. As a result the multilayer contains column–like structure of stacked QDs. Such structures possess high deformation energy, which can lead to the formation of defects. On the other hand, these vertically correlated QDs have many advantages in comparison with single-layer QD structures. Most importantly, the size of the QDs increases considerably when going from the bottom to the top layer [3], leading to a red shift of the emission wavelength [4]. At the same time, the dot density and the barrier height are approximately the same in all the layers of the multilayer.

Control of the growth and further optimisation of advanced QD structures necessitates application of adequate diagnostic techniques capable to characterise them quickly and accurately. Optical characterisation methods offer fast, non-destructive and contactless option. The most popular technique, photoluminescence (PL), works best at low temperatures and yields mainly ground-state interband transitions. On the other hand, photomodulated reflectance (PR) spectroscopy [5], which is rarely used experimental method for checking of QD structures [6], uses laser pump source to periodically perturb sample dielectric function. The alternating component of the sample reflectance, which is subsequently measured, exhibits sharp derivative-like spectral features in the region of interband transitions. As a result, PR provides equivalent energy resolution to that of PL at low temperatures and probes wider range of critical points, highlighting the ground-state and also many high-order interband optical transitions from which band structure, especially of the wetting layer, can be deduced.

In this contribution, we present report on photoluminescence and photomodulated reflectance characterisation of various types of MOVPE grown InAs/GaAs QD structures and we show how the PR in combination with PL is useful for the characterisation of QD structures.

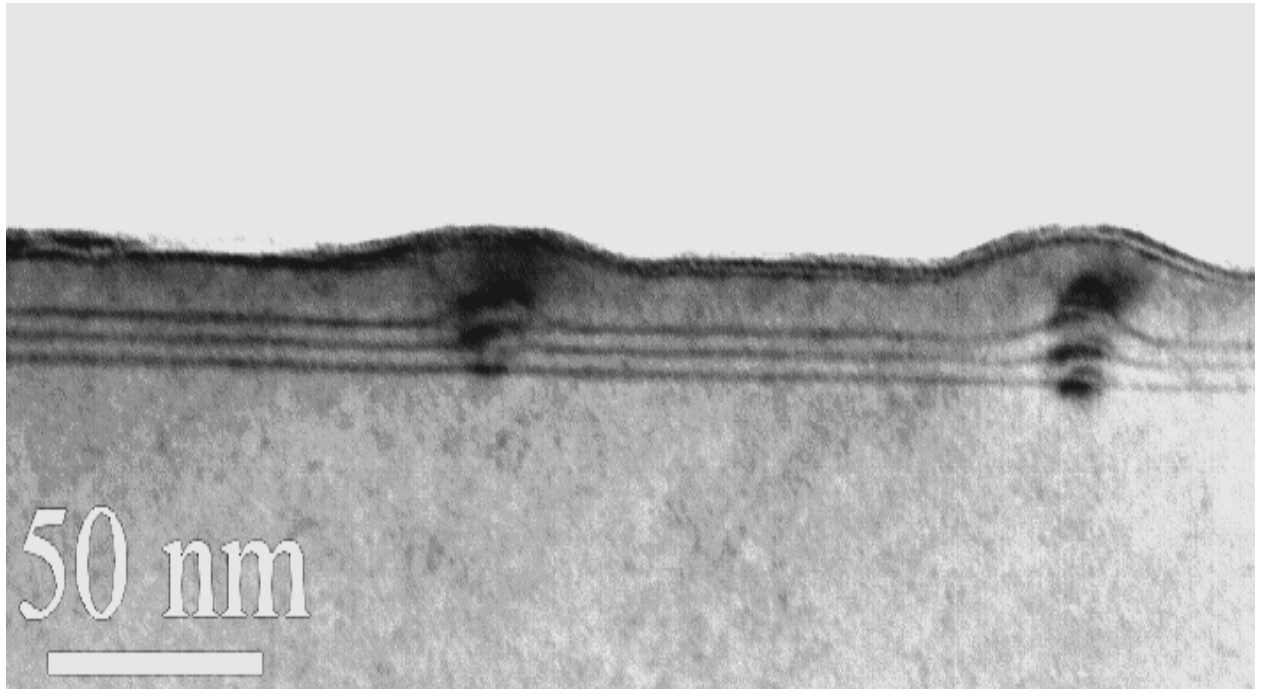

**Fig. 1.** TEM image of the sample with three InAs QD layers separated by 3.7 nm thick layers of GaAs.

## 2. EXPERIMENTAL

Single- and multilayer InAs/GaAs QD structures were prepared by LP MOVPE technique in an AIXTRON 200 machine using Stranski–Krastanow growth mode on semi-insulating GaAs (100) substrates. Precursors used for the growth of GaAs and InAs layers were TMGa, TMIn and $AsH_3$. Prior to the growth, the substrate temperature was increased to 800 °C for 5 min under $AsH_3$ flow. The total pressure of 70 hPa and total flow rate through the reactor of 8 slpm was kept during the growth. The first GaAs buffer layer was grown at 650 °C. Subsequently, the temperature was decreased to 500 °C for the growth of the rest of the structure, consisting of the second GaAs buffer layer, the InAs QD layer and the GaAs capping layer 15 nm thick. We studied the influence of the growth conditions on the QD properties. Namely, the growth interruption time (GIT) was changed from 30 s to 7.5 s for single layer structures and the number of QD layers was changed from 1 to 7.

Due to the high probability of radiative recombination in InAs QD/WLs, photoluminescence (PL) is often used as a convenient tool for the characterisation and optimisation of QD properties. PL of these QD/WL structures was measured at 7 and 300 K. PL was excited by an Ar and a semiconductor lasers (488 nm and 808 nm) with excitation density ~ 5 W $cm^{-2}$. Standard lock in technique with Ge detector was used for detection.

Photomodulated reflectance (PR) spectra were measured using a setup where a 5 mW HeNe laser was used as the modulating source and a 30 W tungsten-halogen lamp filtered by a JY640 grating monochromator provided the monochromatic light. The reflected light was detected by a cooled Ge detector and the resulting signal was processed by PAR5210 lock-in amplifier.

The interpretation of measured optical data was based on simulation of electronic states in InAs and InGaAs layers using the nextnano$^3$ simulator [7]. For a given structure, the computation started by globally minimising the total elastic energy using a conjugate gradient method. This yielded the local strain tensor which in turn determined the band offsets and light/heavy holes splitting. Subsequently, the multi-band-Schrödinger and Poisson equations were solved using the GaAs, InAs and InGaAs band parameters from [8].

## 3. RESULTS AND DISCUSSION

The example of studied QD multilayer structure is shown in Fig. 1, where standard transmission electron microscopy (TEM) picture of three QD layers sample is shown. The column–like structure of stacked QDs is clearly seen and the thickness of GaAs separation and capping layers can be determined, however, the thickness of wetting layers (WL) is falsified by strain and there is also a great uncertainty about its thickness. Since optical transitions in ultrathin InAs quantum well structures are highly sensitive to the thickness and In content in the wetting layers, we used a combination of optical methods (PL and PR) for quick and accurate characterisation of grown structures.

First we studied the effect of growth interruption time change on QD properties. The growth interruption after the InAs layer growth, which is needed to complete QDs formation, was changed in the range from 7.5 to 30 seconds. PL spectra measured at 7 K of samples with different GIT are shown in Fig. 2. It is clearly shown that emission from QD is shifted towards lower energy with increasing GIT. The PL red shift can be explained by increase of QD size with increasing GIT. These results show that QD formation is very slow process. It is also obvious that the thickness of the WL changes, however, lack of PL signal from the WL does not allow to prove it. To get more information about WL, we monitored optical transitions by PR measurements. Spectra received by PR spectroscopy on samples with QD structures grown with different GIT are shown in Fig. 3. Spectra are dominated by GaAs band gap bulk-like signal at the energy of 1.42 eV. Below this energy the spectrum exhibits a broad resonance in the band of QDs transitions and sharp resonances corresponding to the optical transitions in the WL. The arrows indicate the transition energies obtained from fitting the PR data with the first derivatives of the Gaussian line shape, which is the appropriate form for confined transitions at room temperature investigated by PR [5].

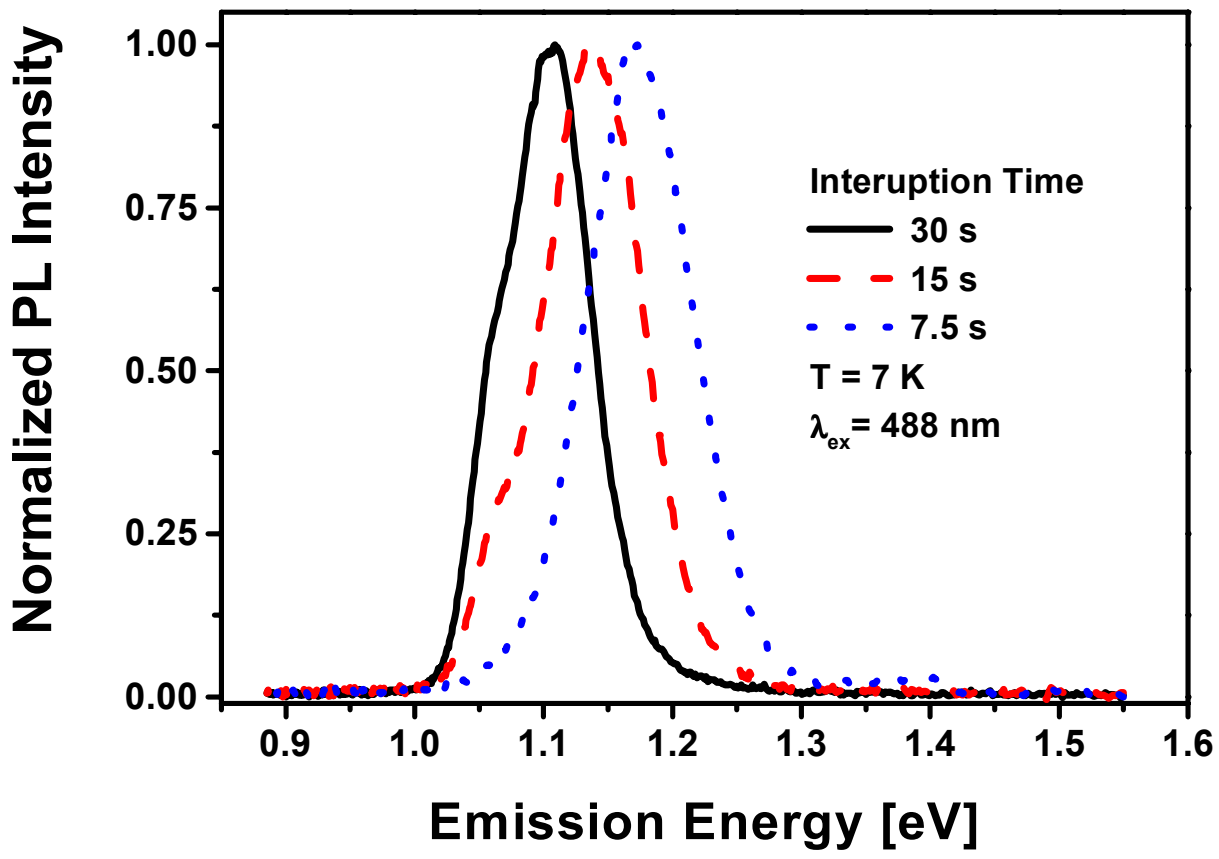

**Fig. 2.** Normalised PL spectra of single QD structures prepared with different growth interruption time measured at 7 K. TEM image of three InAs QD layers sample with 3.7 nm thick separation GaAs layer.

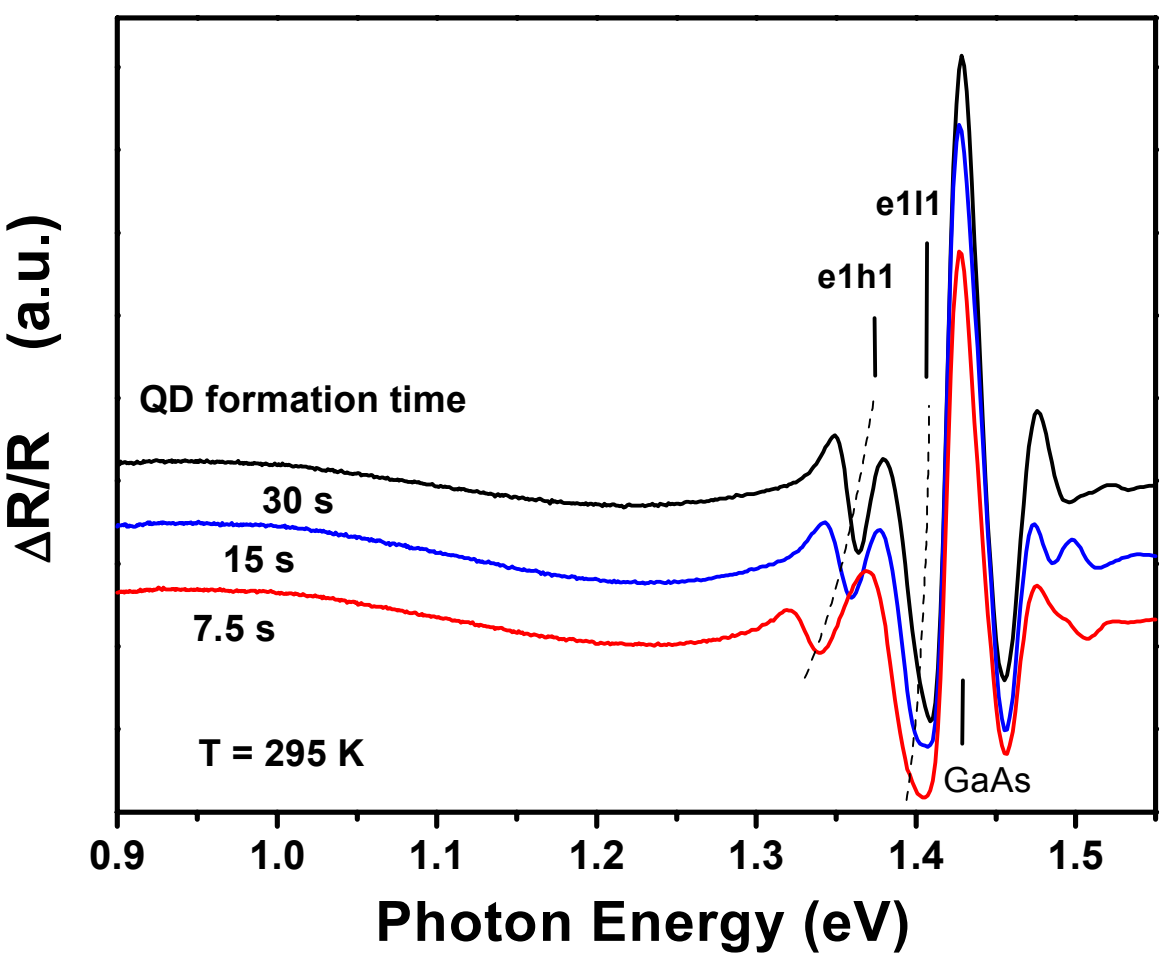

**Fig. 3.** Room temperature photoreflectance spectra of samples with a single QD layer where different growth interruptions were used for QDs formation. Theoretical quantum transitions in the wetting layer (WL) identified by simulation are also indicated.

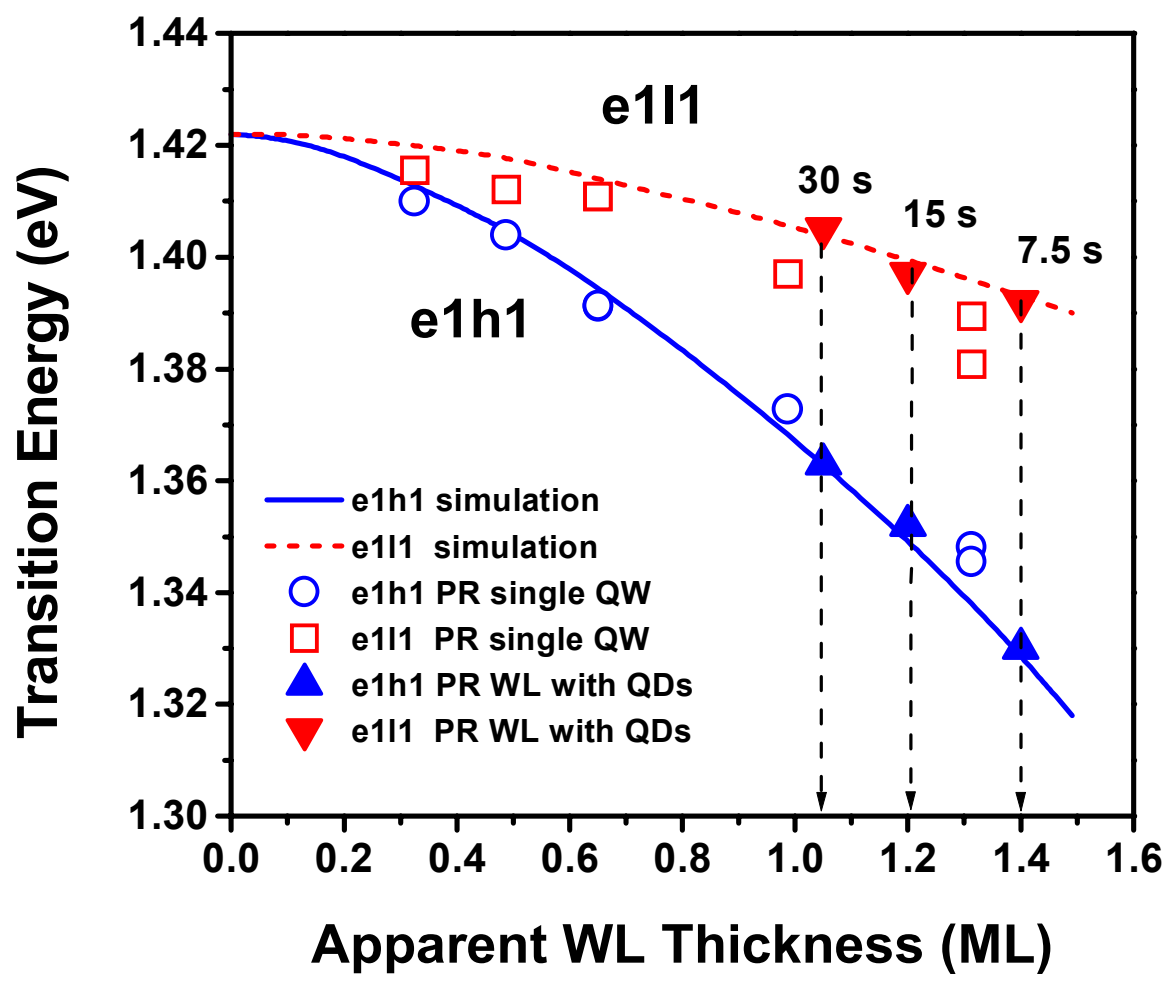

**Fig. 4.** Measured (PR) and calculated (lines) room temperature transition energies for an ultrathin MOVPE grown InAs layer as a function layer thickness (open). Triangles indicate e1h1 and e1l1 transitions identified in wetting layers of QD structures grown with different GIT.

Particular transitions were identified by electronic state simulation. The strongest resonance at lowest energy is attributed to e1h1 transition (the transition between the fundamental states in conduction and heavy hole subbands). Besides this, the PR spectrum shows weaker e1l1 transition (the lowest energy transition for light holes). Fig. 3 clearly shows that increasing of the GIT causes blue shift and lower splitting of e1h1 and e1l1 transitions. This means attenuation of WL thickness. To get exact magnitudes of WL thickness, we compared the measured e1h1 and e1l1 transitions from Fig. 3 with those received on ultrathin InAs quantum wells (QW) grown by the same apparatus [10]. This is shown in Fig. 4 together with the theoretical splitting received from simulation. Fig. 4. clearly shows that increasing of the GIT substantially decreases the apparent WL thickness (from close-to-relaxation limit of 1.4 ML for GIT=7.5 s to 1.05 ML for GIT=30 s) and corresponding strain in the structure. At this point, it is worth to mention, that our MOVPE grown ultrathin (~1 ML) InAs layers are in fact formed by ~2.5 ML thick InGaAs well where the In atom distribution can be assumed to be Gaussian [10].

The results of PL measurement of samples with different number of vertically stacked InAs QD layers are shown in Fig. 5 where normalised PL spectra measured at room temperature are presented. The intensity of PL structure with 5 QD layer was very low and the half width at full maximum (HWFM) ~ 100 meV show on poor quality of prepared structure. The preparation conditions were optimised on the structure with 3 InAs QD layers. Position of PL maxima of 3 QD layer sample is shifted towards low energy in comparison with the single layer sample. This PL red shift is caused by increase of QD size with increasing of layer number. Increasing the number of QD layers also causes coupling of QW formed by their WLs. Again, results of PR measurement on these structures, which are shown in Fig. 6, provided valuable information even for samples with low luminescence intensity (higher number of QDs layers). Fig. 6 shows that increasing number of QD's layers decreases broad resonance in the QD's energy band (0.9 to 1.1 eV) and causes a red shift of all resonances originating from the WLs. Simulation results show that this is due to the coupling of the electronic states in QWs formed by WLs which causes splitting and subsequent mixing of the confined twofold-degenerate electron, hh and lh-states of a single well into a pair of symmetric and antisymmetric states. As a result, red shift of the e1h1 and e1l1 levels and appearance of the mixed e2h1 transition is observed. Although e2h1 transition is forbidden, the built-in electric field in the structure makes it observable in PR measurement. Close agreement between experiment and simulation for the magnitude of GaAs separation layer thickness of 7.5 nm, which was confirmed XRD measurement, indicate on the validity of the model which was used for simulation of multi QWs formed by WLs [10]. It is worth to point out, that the simulation results are very sensitive to WL separation thickness, the In content and the spreading of the In profile in the WL. In this way, the PR measurement supported by calibrated simulation can provide valuable information about growth and optimisation of vertically stacked QDs layers.

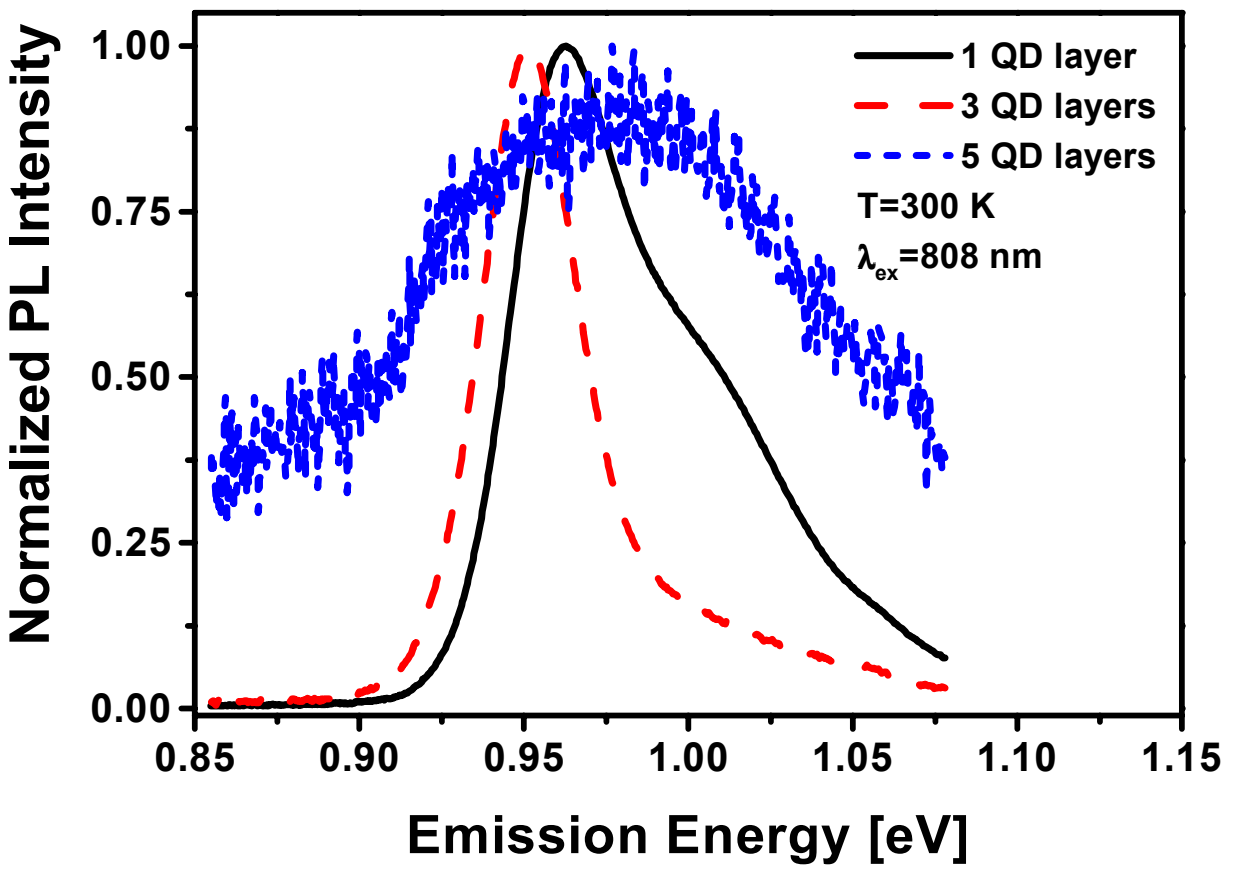

**Fig. 5.** Room temperature photoluminescence spectra of single layer of QDs structures and multiple-layer of vertically correlated QDs structures separated by 7.5 nm thick GaAs layers.

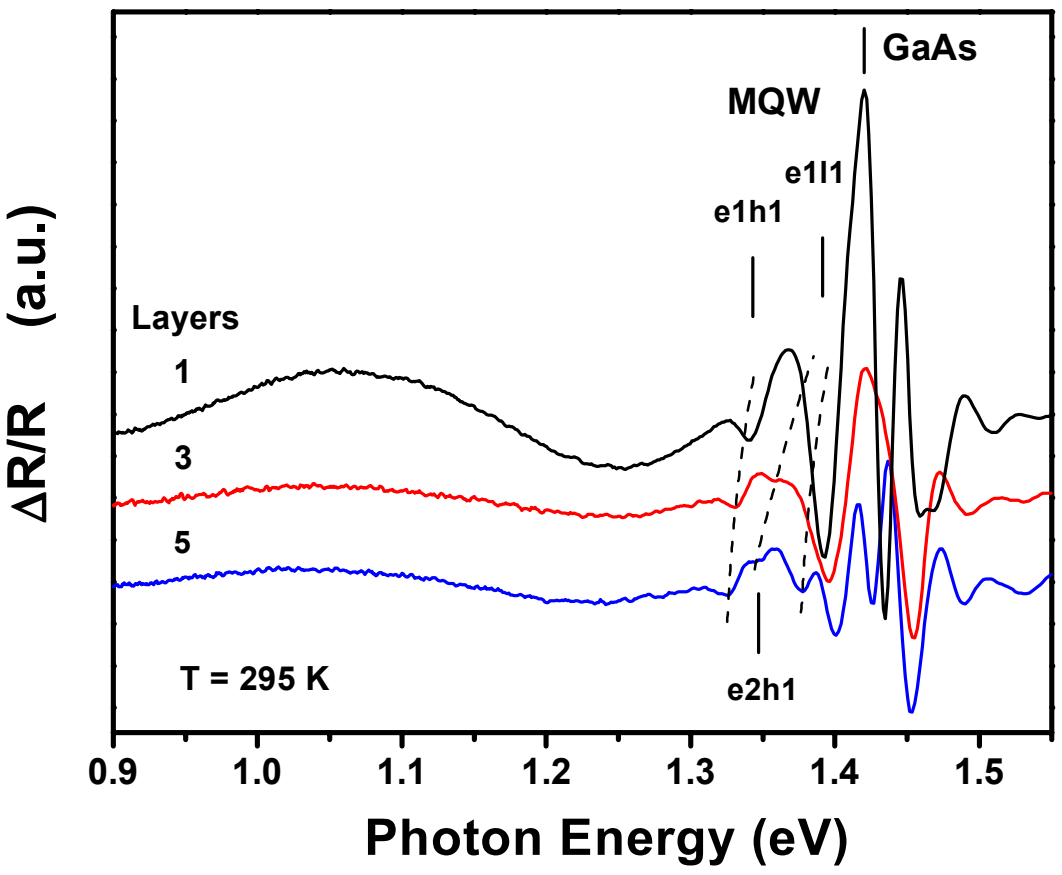

**Fig. 6.** Room temperature photoreflectance spectra of samples with 1, 3 and 5 layers of InAs quantum dots. Theoretical quantum transitions in the multiple QW structure consisting of wetting layers identified by simulation are also indicated.

## 4. CONCLUSIONS

We have shown that photomodulated reflectance spectroscopy in combination photoluminescence can be used for quick and accurate characterisation of different types of quantum dot structures. Results show that photoreflectance spectroscopy is an excellent tool for characterisation of wetting layers (thickness and composition) and for spacer analysis of vertically stacked QDs structures. The techniques is very useful especially in the case of structures with low luminescence intensity.

## 5. ACKNOWLEDGEMENTS

This work was supported by the grant No. 202/06/0718 of the Grant Agency of the Czech Republic, by the Research Programme MSM 6840770014, the GAAV grant No. B101630601, and the project AV0Z 10100521.